\documentclass[twocolumn,prd,aps,amsmath,amssymb,nofootinbib]{revtex4-1}
\usepackage{slashed}
\usepackage{amsmath}
\usepackage{epsfig}
\usepackage{amsfonts}
\usepackage{amssymb}
\usepackage{bm}

\begin{document}
\title{Repulsive to Attractive Fluctuation-Induced Forces\\
in
Disordered Landau-Ginzburg Model}
\author{C. D. Rodr\'iguez-Camargo}
\email{email address: christian.rodriguez-camargo.21@ucl.ac.uk }
\author{A. Saldivar}
\email{email address: asaldivar@cbpf.br}
\author{N.~F.~Svaiter}
\email{email address: nfuxsvai@cbpf.br}
\affiliation{Department of Physics and Astronomy, University College London, London WC1E 6BT, United Kingdom, }
\affiliation{Centro Brasileiro de Pesquisas F\'{\i}sicas, 22290-180 Rio de Janeiro, RJ, Brazil}

\begin{abstract}

Critical fluctuations of some order parameter describing a fluid generates long-range forces between boundaries. 
Here, we discuss fluctuation-induced forces associated to a disordered Landau-Ginzburg model defined in a $d$-dimensional slab geometry  $\mathbb R^{d-1}\times[0,L]$.  In the model the strength of the disordered field is defined by a non-thermal control parameter. We study a nearly critical scenario, using the distributional zeta-function method, where the quenched free energy is written as a series of the moments of the partition function.
In the Gaussian approximation, we show that, for each moment of the partition function, and for some specific strength of the disorder, the non-thermal fluctuations, associated to an order parameter-like quantity, becomes
long-ranged. We demonstrate that the sign of the fluctuation induced force between boundaries, depend in a non-trivial way on the strength of the aforementioned non-thermal control parameter. 
\end{abstract}

\maketitle
\section{Introduction}\label{intro}

Fluctuations-induced forces are a strikingly universal phenomena since macroscopic boundaries that changes the spectrum of a fluctuating medium may present such type of associated forces. Examples of this phenomena are the Casimir forces generated by quantum fluctuations \cite{casimir,ca7,ca2,ff,new,new2}. Situations as a bounded medium experiencing thermal fluctuations near to a critical regime with long range correlations, or Goldstone modes of a broken continuous symmetry, may lead to the appearance of fluctuation-induced forces 
\cite{cce16,krech,cce19,bran,cce18,dohm,17}.  

Critical regimes are also achieved in fluids and magnetic systems with quenched disordered fields \cite{nattermann1,nattermann}. Considering this scenario, and inspired in the critical Casimir effect, in this work, we study the associated induced force that appears in a system described by disordered Landau-Ginzburg model defined in a $d$-dimensional slab geometry, which is driven to the criticality by non-thermal fluctuations. In a confined system approaching a second-order phase transition, when the length scale of the fluctuations are very large, an influence from the boundaries may appear. The fluctuation spectrum associated to the order parameters-like quantities, becomes highly sensitive to the geometry of the boundaries.  The terminology order parameter-like quantities will be discussed later.

In order to deal with critical regimes droved by quenched disorder fields, we employ the distributional zeta-function method. This methods leads to a representation for the quenched free energy where the main contribution is given by a series. Each term of this series is a moment of the partition function with its own ground state. Therefore the multivalley free energy landscape of some disordered systems can be easily obtained \cite{distributional,distributional2,zarro1,zarro2,bose-einstein,polymer,haw,spin-glass}.  

Our purpose is to discuss the sign of the force between the boundaries, for the case of Dirichlet boundary conditions in the nearly critical scenario. To proceed, in each moment of the partition function we compute the saddle-point contribution and discuss Gaussian fluctuations around such saddle-points. Next, we deal with the series of the eigenvalues of Laplace operators. Using generalized zeta-functions, and an analytic regularization procedure, we develop a global approach following the Ref. \cite{ca11}. This procedure shows that there are specific moments of the partition function which are contributing to the force between the boundaries, induced by geometric restrictions, i.e., the constraints in the fluctuation spectra  of each specific moment. Although this global approach does not show the connection between the structure of the divergences and the geometry of the boundaries, its simplicity reveals the relation between the intensity of the effect and the correlation lengths of the fluctuations associated to the order parameters-like quantities in some moments of the partition function. Also, it shows the link among the dimension of the space, the boundary conditions, and the structure of the divergences in the associated spectral zeta-functions.

Repulsive and attractive critical Casimir forces depending on the boundary conditions was discussed in Ref. \cite{bonn}. In Ref. \cite{diehl2}, for systems described by an $O(N)$ $\varphi^{4}$ model in a $d$-dimensional film geomery, it was proved that there is a crossover from attractive to repulsive induced forces, as a function of the distance between the boundaries. For quantum fields, a similar result can be found in Ref. \cite{atrac1}, where it was discussed the sign of the Casimir force between two plates, a perfectly conducting one and an infinite permeable plate. In the Ref. \cite{atrac2} it was proved that a repulsive Casimir force appears when the boundaries are dielectric materials with nontrivial magnetic susceptibility. 
%See also the Ref. \cite{farina}. 
Finally, in 
Refs. \cite{caruso,asorey} it was discussed the dependence of the sign of the Casimir energy on the dimensionality of space, type of boundary condition and others variables. Our main result is a connection between the sign of the fluctuations induced force and the strength of the non-thermal control parameter. The result that the fluctuation induced forces, attractive or repulsive, depends on  the strength of the disorder, as far as we know is new in the literature.

Note that although we are in the statistical field theory framework, we are not using an ultraviotet cut-off in the model. Using the argument of universality in the critical behaviour, where the results of macroscopic measurements must be independent of the cut-off parameter, we can remove a natural physical cut-off and use an analytic regularization procedure to obtain finite results. Although these two methods, the cut-off method and analytic regularization procedures, are quite different in its grounds, it is possible to compare them and prove the analytic equivalence between them in some specific situations \cite{ca12,ca13,ca14,ca15}. One comment is in order. To implement the renormalization program in systems where translational invariance is broken it is required the introduction of counterterms which are surface interactions
\cite{zi,z12,boundary1,boundary2,boundary3}. Since in this work we adopt a global approach, we are not introducing these boundary contributions in the model.

This paper is organized as follows. In section \ref{sec:randommass} we discuss  the Landau-Ginzburg model defined in the continuum, in the presence of a quenched disorder, and the distributional zeta-function method. 
In section \ref{sec:FracLangevinDyn}, in this scenario of confined fluctuations near the critical regime, the spectral zeta-function method and an analytic regularization procedure is discussed.  
Conclusions are given in section \ref{sec:con}. 
Here we are using that $\hbar=k_{B}=1$.

\section{Landau-Ginzburg model with disordered fields} \label{sec:randommass}
\iffalse
In the physics of disordered systems, the random field Ising model is under intensive theoretical, experimental and numerical studies \cite{Fytas1,Fytas2}. This model was introduced by Larkin, to study vortices in superconductors \cite{larkin}. Applying a uniform external field in a diluted Ising antiferromagnet, the random field Ising model can be produced in the laboratory \cite{anti}. Its continuous version is a Landau-Ginzburg model with a random field linearly coupled with the coarse grained field.
\fi
We discuss a confined random field fluid system  assuming a Landau-Ginzburg model with $Z_{2}$ symmetry, in a $d$-dimensional slab geometry $\mathbb R^{d-1}\times [0,L]$. The quenched disorder field is linearly coupled with a scalar field. The cases of the Dirichlet, Neumann Laplacian and also periodic boundary conditions are discussed. In the statistical field theory scenario,  the action functional $S(\varphi)$ for the one component scalar field is given by  
\begin{equation}
S=\int d^{d}\textbf{x}\left[\frac{1}{2}\varphi(\textbf{x})\Bigl(-\Delta+m_{0}^{2}\Bigr)\varphi(\textbf{x})+\frac{\lambda_{0}}{4!}\varphi^{4}(\textbf{x}) \right].
\end{equation}

\noindent The symbol $\Delta$ denotes the Laplacian in 
$\mathbb{R}^{d}$ and $\lambda_{0}$ and $m_{0}^{2}$ are  respectively the coupling constant and a parameter that gives the distance of the model from the critical point. We call it the square mass of the model. Note that we are using the action $S(\varphi)=\beta H(\varphi)$, where $H(\varphi)$ is the Hamiltonian of the model. The action is the energy measured in units of temperature. The generating functional of correlation functions for one disorder realization in the presence of an external source $j(\textbf{x})$, is defined as
\begin{equation}
Z(j,h)=\int[d\varphi]\,\, \exp\left(-S(\varphi,h)+\int d^{d}x\, j(\textbf{x})\varphi(\textbf{x})\right),
\label{eq:disorderedgeneratingfunctional}
\end{equation}
where $[d\varphi]$ is a formal Lebesgue measure, given by $[d\varphi]=\prod_{\textbf{x}} d\varphi(\textbf{x})$ and the action functional in the presence of disorder is 
\begin{equation}
S(\varphi,h)=S(\varphi)+ \int d^{d}x\,h(\textbf{x})\varphi(\textbf{x})
\end{equation}
for $h(\textbf{x}) \in$ $L^{2}(\mathbb{R}^{n})$.
In the above equation $S(\varphi)$ is the pure Landau-Ginzburg action functional, and $h(\textbf{x})$ is a quenched random field. This is the simplest scalar model with  a disorder field linearly coupled to the scalar field of the theory. We would like to point out that one can discuss also the quenched random mass model given by
\begin{equation}
S(\varphi,\eta)=S(\varphi)+\frac{\rho}{4}
\int d^{d}x \,
\eta(\textbf{x})\varphi^{2}(\textbf{x}).\label{eq:spe1}
\end{equation}
This model is known as the random-temperature disorder, where small density of impurities lead to randomness in the local transition temperature.
In this work we will discuss only the quenched random field model. 
Measured in units of temperature,
the disordered free energy for one disorder realization is $W(j,h)=-\ln Z(j,h)$. Performing the average over the ensemble of all realizations of the disorder we have 
\begin{equation}
\mathbb{E}\bigl[W(j,h)\bigr]=\int\,[dh]P(h)\ln Z(j,h),
\end{equation}
where  $[dh]=\prod_{x} dh(x)$ is a functional measure. The probability distribution of the disorder is written as $[dh]\,P(h)$, being
\begin{equation}
P(h)=p_{0}\,\exp\bigl(-\frac{1}{2\,\sigma^{2}}\int\,d^{d}x\bigl(h(\textbf{x})\bigr)^{2}\bigr) .
\end{equation}

The quantity $\sigma$ is a positive parameter associated with the disorder and $p_{0}$ is a normalization constant. 
This defines a delta correlated process 
\begin{equation}
\mathbb{E}[{h(\textbf{x})h(\textbf{y})}]=\sigma^{2}\delta^{d}(\textbf{x}-\textbf{y}) .
\end{equation}

For a given probability distribution of the disorder, one is mainly interested in obtaining the average free energy. For a general disorder probability distribution, using the disordered functional integral $Z(j,h)$ given by Eq. \eqref{eq:disorderedgeneratingfunctional}, the distributional zeta-function, $\Phi(s,j)$, is defined as
\begin{equation}
\Phi(s,j)=\int [dh]P(h)\frac{1}{Z(j,h)^{s}},
\label{pro1}
\vspace{.2cm}
\end{equation}
\noindent for $s\in \mathbb{C}$, this function being defined in the region where the above integral converges. The average generating functional can be written as 
\begin{equation}
\mathbb{E}\bigl[W(j,h)\bigr]=-(d/ds)\Phi(s,j)|_{s=0^{+}}, \,\,\,\,\,\,\,\,\,\, \Re(s) \geq 0,  
\end{equation}
where one defines the complex exponential $n^{-s}=\exp(-s\log n)$, with $\log n\in\mathbb{R}$.
Using analytic tools, again in units of temperature the quenched free energy of a system in the presence of an external field  is $F_{q}(j)=-\mathbb{E}\bigl[W(j,h)\bigr]$,
where 
\begin{align}
&\mathbb{E}\bigl[W(j,h)\bigr]=\sum_{k=1}^{\infty} \frac{(-1)^{k+1}a^{k}}{k k!}\,\mathbb{E}\,\bigl[\bigl(Z(j,h)\bigr)^{\,k}\bigr]\nonumber  \\
&-\ln(a)+\gamma+R(a,j).
\label{m23e}
\end{align}
The quantity $a$ is a dimensionless arbitrary constant, $\gamma$ is the Euler-Mascheroni constant, and $R(a)$ is given by
\begin{equation}
R(a,j)=-\int [dh]P(h)\int_{a}^{\infty}\,\dfrac{dt}{t}\, \exp\Bigl(-Z(j,h)t\Bigr).
\end{equation} 
Integrating over the disorder, each moment of the partition partition function can be written as 
\begin{equation}
\mathbb{E}\,\bigl[(Z(j,h))^{k}\bigr]=\int\,\prod_{i=1}^{k}[d\varphi_{i}]\,\exp\bigl(-S_{\textrm{eff}}(\varphi_{i},j_{i})\bigr),
\end{equation} 
where the effective action $S_{\textrm{eff}}(\varphi_{i})$  describes a $k$-field component field theory. 
\begin{eqnarray}
&&S_{\textrm{eff}}(\varphi_{i}^{(k)},j_{i}^{(k)})=\nonumber \\
&&\int d^{\,d}x\Biggl[\sum_{i=1}^{k}\biggl(\frac{1}{2}\varphi_{i}^{(k)}(\textbf{x})\bigl(-\Delta+m_{0}^{2}\bigr)\varphi_{i}^{(k)}(\textbf{x})\nonumber\\
&&+\frac{\lambda_{0}}{4!}\bigl(\varphi_{i}^{(k)}(\textbf{x})\bigr)^{4}\biggr)-\frac{\sigma^{2}}{2}\sum_{i,j=1}^{k}\varphi_{i}^{(k)}(\textbf{x})\varphi_{j}^{(k)}(\textbf{x})\nonumber \\
&&-\sum_{i=1}^{k}\varphi_{i}^{(k)}(\textbf{x})j_{i}^{(k)}(\textbf{x})\Biggr].
\label{Seff1}
\end{eqnarray}
In order to avoid unnecessary complications, and for practical purposes, we assume the following configuration of the scalar fields $\varphi^{(k)}_{i}(x)=\varphi^{(k)}_{j}(\textbf{x})=\varphi^{(k)}(\textbf{x})$ in the function space and also $j_{i}^{(k)}(\textbf{x})=j_{l}^{(k)}(\textbf{x})=j^{(k)}(\textbf{x})$. All the terms of the series have the same structure and one minimizes each term of the series one by one. 
Assuming that in each moment of the partition function the fields are equal, we have that the $k$-th moment of the partition function is written as 
\begin{equation}
\mathbb{E}\,\bigl[\bigl(Z(j,h)\bigr)^{k}\,\bigr]=\Bigl[\int\,[d\varphi]\,\exp\Bigl(-S^{(k)}%_{\textrm{eff}}
(\varphi^{(k)},j^{(k)})\Bigr)\Bigr]^{k}
\end{equation} 
In this case, the new effective action is written as:
\begin{eqnarray}
&&S^{(k)}(\varphi^{(k)},j^{(k)})=\nonumber \\
&&\int d^{\,d}x\biggl(\frac{1}{2}\varphi^{(k)}(\textbf{x})\Bigl(-\Delta+m_{0}^{2}-k\sigma^{2}\Bigr)\varphi^{(k)}(\textbf{x})\nonumber \\
&&+\frac{\lambda_{0}}{4!}\bigl(\varphi^{(k)}(\textbf{x})\bigr)^{4}-\varphi^{(k)}(x)j^{(k)}(\textbf{x})\biggr).
\label{Seff2}
\end{eqnarray}
In what follows we define each contribution  $W^{(k)}(j)$ as
\begin{equation}
W^{(k)}(j)\equiv c_{k}\,\mathbb{E}\bigl[\bigl(Z(j,h)\bigr)^{k}\,\bigr]
\end{equation}
being $c_{k}(a)=(-a)^{k+1}/kk!$. For simplicity we shall adopt the convention that $c_{k}(a)=c_{k}$.

In this situation, we have three different contribution of the terms of the series depending on the sign of $m_{0}^{2}-k\sigma^{2}$ for $m_{0}^{2}>\sigma^{2}$: \textbf{(i)} the case where $m_{0}^{2}-k\sigma^{2}>0$, 
\textbf{(ii)} the case where $m_{0}^{2}\cong k\sigma^{2}$, is a situation similar to a second-order phase transitions, defines $k_c$
\textbf{(iii)} when this quantity is negative, one has to shift field to a new minimum, i.e., $\phi^{(k)}(\textbf{x})=\varphi^{(k)}(\textbf{x})\pm\vartheta^{(k)}$, with $\vartheta^{(k)}=\left(6(k\sigma^{2}-m_{0}^{2})/\lambda_{0}\right)^{1/2}$. This case is similar to the spontaneous symmetry breaking in statistical field theory and the behaviour of each term of the series is described by the $(+)$ or $(-)$ cases. From now on, we choose the minus sign above. In this case, we find a positive squared mass with self-interactions terms $(\phi^{(k)}(\textbf{x}))^{3}$ and $(\phi^{(k)}(\textbf{x}))^{4}$.
\iffalse
 with the potential $V(\phi^{(k)})=\bigl(k\sigma^{2}-m_{0}^{2}\bigr)\bigl(\phi^{(k)}(\textbf{x})\bigr)^{2}+
\sqrt{\frac{\lambda_{0}}{3!}}\bigl(k\sigma^{2}-m_{0}^{2}\bigr)\bigl(\phi^{(k)}(\textbf{x})\bigr)^{3}
+\frac{\lambda_{0}}{4!}\bigl(\phi^{(k)}(\textbf{x})\bigr)^{4}$. 
\fi

Here, we consider the three level approximation. In each moment of the partition function order parameters-like quantities are defined, i.e., $\varphi_{0}^{(k)}(\textbf{x})$. In this way, for any moment of the partition function this three level contribution is
\begin{equation}
-\Delta\varphi_{0}^{(k)}(\textbf{x})+\bigl(m_{0}^{2}-k\sigma^{2}\bigr)\varphi_{0}^{(k)}(\textbf{x})+\frac{\lambda_{0}}{3!}\bigl(\varphi_{0}^{(k)}(\textbf{x})\bigr)^{3}=j^{(k)}(\textbf{x}).
\end{equation} 
The Fourier transform of the susceptibility-like quantity is 
\begin{equation}
\chi^{(k)}(\bold{q})=\frac{1}{\bold{q}^{2}+m_{0}^{2}-k\sigma^{2}+\frac{1}{2}\lambda_{0}\bigl(\varphi^{(k)}_{0}\bigr)^{2}}.
\end{equation}

When we have the terms in the series where $k<k_{c}$, we obtain that $\varphi^{(k)}_{0}(\textbf{x})=0$ and
\begin{equation}
\chi^{(k)}(\bold{q})=\frac{1}{\bold{q}^{2}+m_{0}^{2}-k\sigma^{2}} .
\end{equation}

The correlation length for $k<k_{c}$ is therefore,
\begin{equation}
\xi^{(k)}_{\,<}(\sigma,m_{0})=(m_{0}^{2}-k\sigma^{2})^{-\frac{1}{2}} .
\end{equation}

For the terms in the series with $k>k_{c}$ we have

\begin{equation}
\bigl(\varphi^{(k)}_{0}\bigr)^{2}=6(k\sigma^{2}-m_{0}^{2})/\lambda_{0}.
\end{equation}

The Fourier transform of the susceptibility-like quantity is
\begin{equation}
\chi^{(k)}(\bold{q})=\frac{1}{\bold{q}^{2}+2\bigl(k\sigma^{2}-m_{0}^{2}\bigr)}.
\end{equation}

Then the correlation length, when $k>k_{c}$, reads as 
\begin{equation}
\xi^{(k)}_{\,>}=\bigl(2(k\sigma^{2}-m_{0}^{2})\bigr)^{-\frac{1}{2}} .
\end{equation}

Note that we are computing the saddle-point contribution and we will take into account Gaussian fluctuations around such saddle-point. Although the critical exponents using this approximation are not correct for dimensions bellow the critical dimension, here we are interested to compute the fluctuation induced force between the boundaries. Radiative corrections are negligible in this scenario. %% REFERENCES %%

\section{Fluctuation-induced force in Systems with Disorder} \label{sec:FracLangevinDyn}

In this section we will discuss the nearly critical scenario of the system, in order to present the fluctuation induced force between the boundaries. The next step is to assume the Gaussian approximation. For $k\sigma^{2}>m_{0}^{2}$ we expand each functional integral around the minimum up to the lowest-order quadratic term and integrate out the fluctuations. 

Starting from the elliptic operator $-\Delta+2(k\sigma^{2}-m_{0}^{2})$ we define,
\begin{equation}
D(\textbf{x},\textbf{y},k)\equiv \Bigl(-\Delta+2\bigl(k\sigma^{2}-m_{0}^{2}\bigr)\Bigr)\delta^{d}(\textbf{x}-\textbf{y}) .
\label{dkxy}
\end{equation}

Within (\ref{dkxy}) we define the inverse kernel $K(\textbf{x},\textbf{z};k)$ as
\begin{equation}
\int d^{d}z\,K(\textbf{x},\textbf{z};k)D(\textbf{z},\textbf{y};k)=\delta^{d}(\textbf{x}-\textbf{y}) .
\end{equation}

Therefore up to the Gaussian approximation we can write 
\begin{align}
&\mathbb{E}[W(j,h)]=\sum_{k=1}^{\infty}\frac{c_{k}}{\bigl(\det D(k)\bigr)^{k/2}}\nonumber \\
&\left[\exp\Biggl(-\int d^{d}x\int d^{d}y\,j^{(k)}(\textbf{x})K(\textbf{x},\textbf{y};k)j^{(k)}(\textbf{y})\Biggr)\right]^{k}.
\end{align}

With the theory in finite-size geometry in one dimension, we have the spatial coordinate $x_{d}=z$ compactified, and a slab defined as

\begin{equation*}
\Omega=[\textbf{x}\equiv (x_{1},x_{2},..,x_{d-1},z):0\leq z\leq L] \subset \mathbb{R}^{d} .
\end{equation*}

In systems where the translational invariance is broken, in a local approach, one can use a Fourier representation for the fields. Since the system possesses translational invariance along the direction parallel to the plates, one has to adopt a mixed representation, to implement the renormalization program. 

The Fourier transform of the susceptibility-like quantity  $\chi^{(k)}((\textbf{x}-\textbf{y})_{||},z,z')$ reads 
\begin{equation}
\chi^{(k)}(\bold{q}_{||},n)=\frac{1}{\bigl(\bold{q}_{||}\bigr)^{2}+
\bigl(\frac{2n\pi}{L}\bigr)^{2}+2\bigl(k\sigma^{2}-m_{0}^{2}\bigr)} .
\end{equation}

Each of moment of the partition function contribute to the quenched free energy by mean of a functional determinant. To evaluate each of these functional determinants the formalism of spectral zeta-function is a standard procedure \cite{plei,seeley,ray,voros1992}. Suppose a infinite sequence of non-zero real or complex numbers $\lambda_{n}$. If the sequence of numbers is zeta regularizable we define the regularized product
$\prod_{n}\,\lambda_{n}$.
The zeta regularized product of these numbers is defined as $\exp\bigl(-\zeta'(0)\bigr)$ where this generalized zeta-function is given by  
\begin{equation}
\zeta(s)=\sum_{n}\lambda_{n}^{\,-s}, \,\,\, \Re\,(s)>s_{0}
\label{zeta10}
\end{equation}
for $s \in \mathbb{C}$, this function being defined in the region of the complex plane where the sum converges and $\zeta'(0)=\frac{d}{ds}\zeta(s)|_{s\rightarrow 0^{+}}$, by analytic extension. In this framework, one can write
\begin{equation}
\Bigl[\text{det}\, D(k)\Bigr]^{-\frac{k}{2}}=\exp\bigl[\frac{k}{2}\bigl(\zeta'(0,k)\bigr)\bigr] .
\end{equation}

Due to the fact that we are assuming flat boundaries, here we will discuss each contribution for the free energy using an analytic regularization procedure, calculating $\zeta(-\frac{1}{2},k)$ instead of  $\zeta'(0,k)$ \cite{blau}. One comment is in order. There is a relationship between the Casimir energy and the one-loop effective action. These two quantities differ by a contribution proportional to the second fundamental form, which is zero for a $d$-dimensional slab geometry. 

Let us assume a \emph{thermodynamic limit} with respect to the surface area, i.e., $L_{1},L_{2},...,L_{d-1}\gg L_{d}$, and  $2\bigl(k\sigma^{2}-m_{0}^{2}\bigr)>0$. 
To proceed one define the spectral zeta-function $\zeta_{\,d}(s,k)$ as
\begin{align}
\zeta&_{\,d}(s,k)=\frac{1}{(2\pi)^{d-1}}\Biggl(\prod_{i=1}^{d-1}L_{i}\Biggr)
\int\,\prod_{i=1}^{d-1}dq_{i}\nonumber\\
&\sum_{n\in \mathbb{Z}}\Biggl(q_{1}^{2}+...+q_{d-1}^{2}
+\biggr(\frac{2\pi n}{L_{d}}\biggl)^{2}+2\bigl(k\sigma^{2}-m_{0}^{2}\bigr)
\Biggr)^{-s}
\end{align}
for $s\in\mathbb{C}$. We like to point out that in order to be rigorous we should have included a term $\mu^{2s+1}$, where $\mu$ have dimension of mass, in the above expression, to keep dimensionality consistence. But in order to avoid caring unnecessary nomenclature and given that we are only looking to the situation were $s=1/2$, we can omitted and stick to our notation. 

With the finite length $L_{d}=L$ and performing the angular part of the integral over the continuous mode spectrum of the $(d-1)$ non-compact dimensions we get that 
\begin{equation}
\int d\Omega_{d-1}=\frac{2\,(\pi)^{\frac{d-1}{2}}}{\Gamma\bigl(\frac{d-1}{2}\bigr)}.
\end{equation}
Note that we are assuming $d\geq 2$.  We would like to stress that one can show that the second-order phase transition in $d=2$ is suppressed, since the two-point correlation functions belongs to the space of locally integrable functions, in the sense of generalized functions, and therefore must have integrable singularities only at coinciding points. Since, $G_{0}\bigl(\textbf{x}-\textbf{y},m^{2}_{0}\,\bigr)=-\frac{1}{2\pi}\ln\,\bigl({m_{0}|\textbf{x}-\textbf{y}|}\bigr)$
in $d=2$ dimensions, the theory violates the regularity condition, that is one condition to define a field theory.
Since the nearly critical scenario is reached when 
the correlation lengths of the fluctuations of the order parameters-like quantities satisfies $\xi^{(k)}_{\,>}> L$. 

Let us define $\lfloor\kappa\rfloor$ as the largest integer $\leq \kappa$ for any $\kappa \in \mathbb{R}$. In other words, $\lfloor\kappa\rfloor$ is the integer $r$ for which $r\leq \kappa<r+1$. Within this notation, we notice that we have a set of moments such that 
\begin{equation}
\lfloor\frac{m^{2}_{0}}{\sigma^{2}}\rfloor\leq k \leq\lfloor\frac{1}{\sigma^{2}}
\bigl(\frac{1}{2L^{2}}+m_{0}^{2}\bigr)\rfloor .
\end{equation}

% where $\lfloor\kappa\rfloor$ denote the largest integer $\leq \kappa$. 
We are interested to discuss the contribution of the moments of the partition function where 
$k\geq \lfloor\frac{m_{0}^{2}}{\sigma^{2}}\rfloor$, i.e., where each of the order parameters-like quantities does not vanish.

At this point we introduced the spectral zeta-function per unit area 
\begin{equation}
Z_{d}(s,k)=\frac{\zeta_{d}(s,k)}{A(d)\bigl(\prod_{i=1}^{d-1}L_{i}\bigr)}
\end{equation}
where the factor $A(d)$ is defined as
\begin{equation}
A(d)=\frac{1}{2^{d-2}\,\pi^{\frac{d-1}{2}}\Gamma\bigl(\frac{d-1}{2}\bigr)}.
\end{equation} 
The expression for $Z_{d}(s,k)$ is written as  
\begin{align}
&Z_{d}(s,k)=\biggl(\frac{L}{{\sqrt{4\pi}}}\biggr)^{2s}\int_{0}^{\infty}dp\,p^{d-2}\nonumber\\
&\sum_{n\in \mathbb{Z}}\Biggl(\pi n^{2}+\frac{L^{2}}{4\pi}\Bigl(p^{2}+2\,\bigl(k\sigma^{2}-m_{0}^{2}\bigr)\Bigr)
%f^{2}(p,k)
\Biggr)^{-s}.
\end{align}
To proceed, let us rewrite $Z_{d}(s,k)$ in a way suitable for our analysis. After a Mellin transform, and renaming some quantities we can rewrite the spectral function per unit area as
\begin{align}
&Z_{d}(s,k)=\frac{B(s,d)}{2\Gamma(\frac{d-1}{2})}\frac{1}{L^{d-2s-1}}
%\biggl(\frac{L\mu}{{\sqrt{4\pi}}}\biggr)^{2s}
\int_{0}^{\infty}dr\,r^{d-2}\nonumber\\
&\int_{0}^{\infty}dt\,t^{s-1}
%\exp^{{-\lambda\bigl(t+\frac{1}{t}\bigr)}}
\exp{\Bigl(-\bigl(m^{2}(k)+r^{2}\bigr)t\Bigr)}
\,\Theta(t),
\label{eqz}
\end{align}
with the dimensionless quantities $m^{2}(k)=\frac{L^{2}}{2\pi}(k\sigma^{2}-m_{0}^{2})$ and $r^{2}=\frac{L^{2}}{4\pi}\,p^{2}$.  Also $B(s,d)$ is defined as
\begin{equation}
B(s,d)=2
(\sqrt{4\pi})^{d-2s-1}\frac{\Gamma(\frac{d-1}{2})}{\Gamma(s)}
\end{equation}
and the theta function $\Theta(v)$ defined as
\begin{equation}
\Theta(v)=\sum_{n\in \mathbb{Z}}\exp\bigl(-\pi\,n^{2}v\bigr)
\end{equation}
an example of a modular form, appears. 
The quantity 
\begin{equation}
m^{2}(k)
=\frac{L^{2}}{4\pi\bigl(\xi_{>}^{(k)}\bigr)^{2}}
\end{equation}
defines the finite size scaling, i.e., close to the critical point, finite size effects are controlled by the ratio $L/\xi_{>}^{(k)}$. Splitting the integral in the $t$ variable from eq. (\ref{eqz}) into two contributions and performing the integral in the $r$ variable we can recast the spectral zeta function per unit area as $Z_{d}(s,k)=Z_{d}^{(1)}(s,k)+Z_{d}^{(2)}(s,k)$. Where
\begin{equation}
Z_{d}^{(1)}(s,k)=C(s,d)\int_{0}^{1}dt\,t^{s-\frac{d}{2}-\frac{1}{2}}
\exp{\bigl(-m^{2}(k)t\bigr)}\Theta(t),
\end{equation}
and
\begin{equation}
Z_{d}^{(2)}(s,k)=C(s,d)\int_{1}^{\infty}dt\,t^{s-\frac{d}{2}-\frac{1}{2}}
\exp{\bigl(-m^{2}(k)t\bigr)}\Theta(t)
\end{equation}
where $C(s,d)=(1/L^{d-2s-1})B(s,d)$.
Changing variables in the integral $Z_{d}^{(1)}(s,k)$ and using the symmetry of the theta-function we have 
\begin{equation}
Z_{d}^{(1)}(s,k)=C(s,d)\int_{1}^{\infty}dt\,t^{-s+\frac{d}{2}-1}
\exp{\bigl(-\frac{m^{2}(k)}{t}\bigr)}\Theta(t).
\end{equation}
From the definition of the psi-function  $\psi(v)=\sum_{n=1}^{\infty}\exp{(-\pi n^{2}v)}$, such that $\psi(v)=\frac{1}{2}\bigl(\Theta(v)-1\bigr)$,
we can rewrite $Z_{d}(s,k)$ as having four contributions, $I_{d}^{(i)}(s,k), i=1,...,4$.  Therefore
\begin{align}
&Z_{d}(s,k)=C(s,d)
\nonumber\\
&\biggl(2I_{d}^{(1)}(s,k)+2I_{d}^{(2)}(s,k)+I_{d}^{(3)}(s,k)+I_{d}^{(4)}(s,k)\biggr),
\end{align}

In order to be more explicit, let us evidence this integrals 
\begin{equation}
I_{d}^{(1)}(s,k)=\int_{1}^{\infty}dt\,t^{s-\frac{d}{2}-\frac{1}{2}}\exp{\bigl(-m^{2}(k)t\bigr)}\psi(t)
\end{equation}

\begin{equation}
 I_{d}^{(2)}(s,k)=\int_{1}^{\infty}dt\,t^{-s+\frac{d}{2}-1}\exp{\biggl(-\frac{m^{2}(k)}{t}\biggr)}\psi(t)
\end{equation}

\begin{equation}
I_{d}^{(3)}(s,k)=\int_{1}^{\infty}dt\,t^{s-\frac{d}{2}-\frac{1}{2}}
%e^{-\lambda\bigl(t+\frac{1}{t}\bigr)}
\exp{\bigl(-m^{2}(k)t\bigr)}
\end{equation}
and finally
\begin{equation}
I_{d}^{(4)}(s,k)=\int_{1}^{\infty}dt\,t^{-s+\frac{d}{2}-1}
%e^{-\lambda\bigl(t+\frac{1}{t}\bigr)}
\exp{\biggl(-\frac{m^{2}(k)}{t}\biggr)}.
\end{equation}
For the case of the Dirichlet Laplacian where only the integrals $I_{d}^{(1)}(s,k)$ and $I_{d}^{(2)}(s,k)$ appears, and using the fact that $\psi(t)=O(e^{-\pi t})$ as $t\rightarrow \infty$, the integrals $I_{d}^{(1)}(s,k)$ and $I_{d}^{(2)}(s,k)$ represent an everywhere regular functions of $s$ for $m^{2}(k)\in \mathbb{R}_{+}$. 
The upper bound insure uniform convergence of the integrals on every bounded domain in $ \mathbb{C}$.
As in the standard quantum field theory scenario, the contribution to the average free energy from each moment of the partition function of the system can be evaluated for $s=-\frac{1}{2}$.

\begin{figure}[h]
	\includegraphics[width=0.54\textwidth]{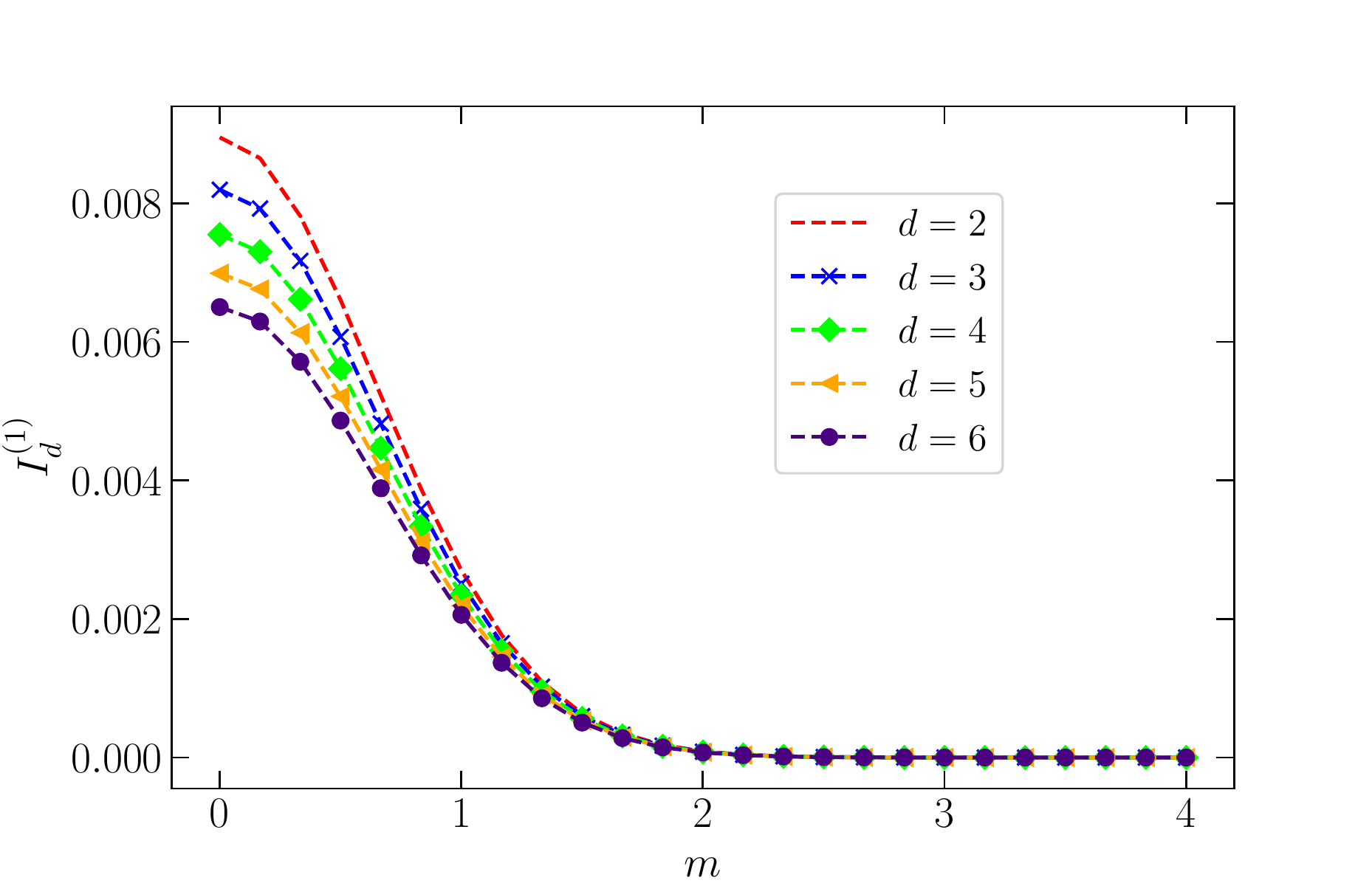}\\
	\caption{Behavior of $I_{d}^{(1)}(s,k)$ for $s=-1/2$, for arbitrary dimensionality of the space and also 
dimensionless quantity $m(k)=m$.}
	\label{fig_cas_1}
 \end{figure}
 
 \begin{figure}[h]
	\includegraphics[width=0.54\textwidth]{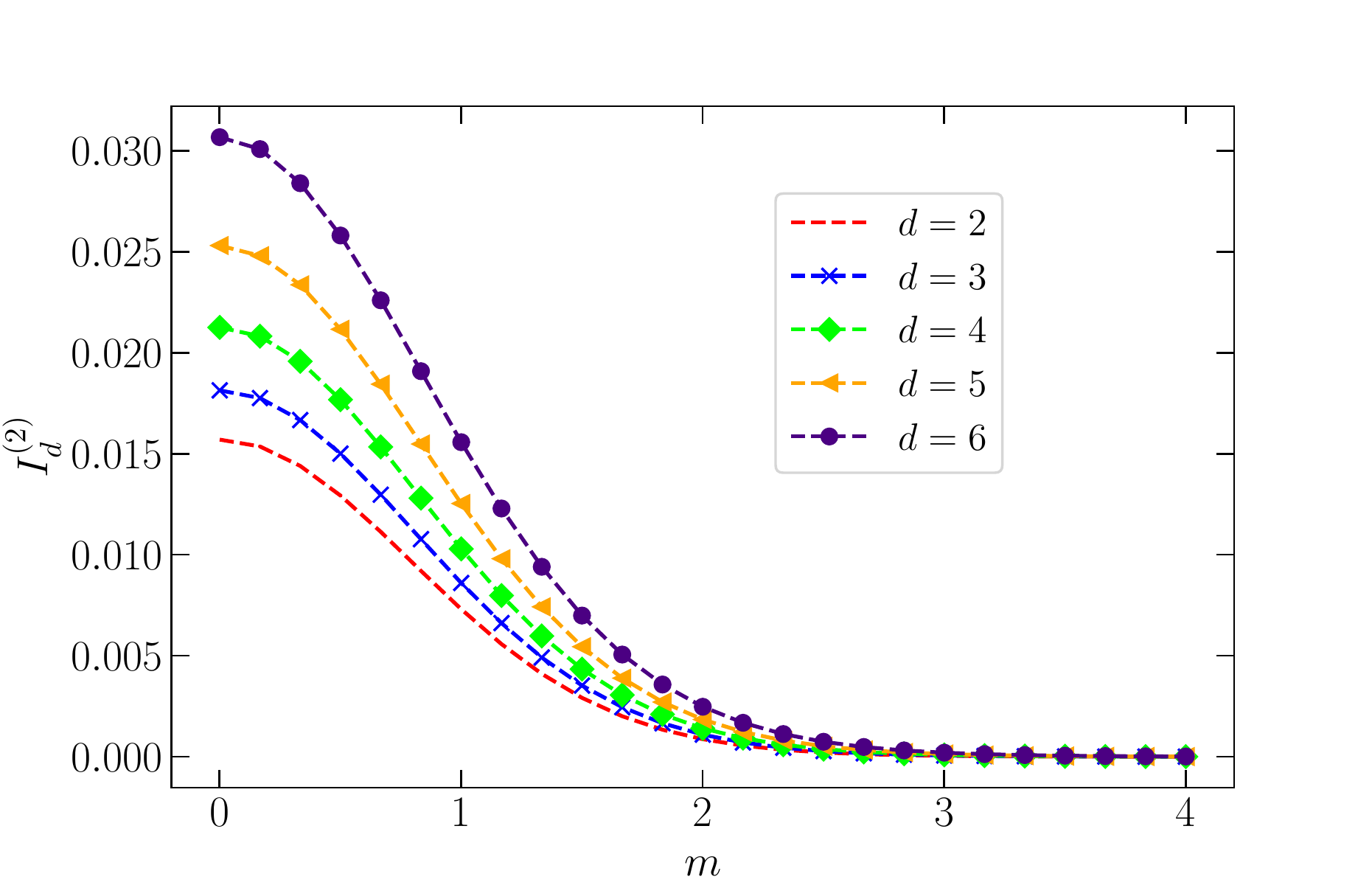}\\
	\caption{Behavior of $I_{d}^{(2)}(s,k)$ for $s=-1/2$, for arbitrary dimensionality of the space and dimensionless quantity $m(k)=m$.}
	\label{fig_cas_2}
 \end{figure}
 In the Fig. (\ref{fig_cas_1}) we depict the behaviour of the integral given by $I_{d}^{(1)}(s,k)$ for an arbitrary dimensionality of space and dimensionless quantity $m(k)$. For completeness  we discuss the $d=2$ case. We can see that the integral vanishes when the value of the dimensionless quantity $m(k)$ satisfies $m(k)>2$. On the other hand the contribution of the integral $I_{d}^{(2)}(s,k)$ is depicted in the Fig. (\ref{fig_cas_2}). The contribution of the integral $I_{d}^{(2)}(s,k)$ for $s=-1/2$ vanishes for $m(k)>2.5$.

For the case of the Neumann Laplacian and also the periodic boundary conditions, not only the integrals $I_{d}^{(1)}(s,k)$ and $I_{d}^{(2)}(s,k)$ but also the integrals $I_{d}^{(3)}(s,k)$ and $I_{d}^{(4)}(s,k)$ also appears. In the absence of the exponential decay of the $\psi(v)$ function and for $m^{2}(k)\in \mathbb{R^{+}}$ we have to discuss the polar structure of the integrals $I_{d}^{(3)}(s,k)$ and $I_{d}^{(4)}(s,k)$. Note that we are assuming that $m^{2}(k)$ is small, but different from zero. One can show that that the contribution of 
$I_{d}^{(3)}(s,k)$ is finite for odd dimensional space. 
Also 
\iffalse

write that the integral $I_{d}^{(3)}(s,k)=\bigl(m^{2}(k)\bigr)^{\frac{d}{2}-s-\frac{1}{2}}\Gamma\biggl(
s-\frac{d}{2}+\frac{1}{2},m^{2}(k)\biggr)$, where $\Gamma(\alpha,x)=\int_{x}^{\infty}dt\,t^{\alpha-1}e^{-t}$ is the incomplete gamma function. Using the series representation for the incomplete gamma function we get
%
\begin{align}
I_{d}^{(3)}(s,k)&=\bigl(m^{2}(k)\bigr)^{\frac{d}{2}-s-\frac{1}{2}}\Gamma\Biggl(s-\frac{d}{2}+\frac{1}{2}\Biggr)\nonumber\\
&+\sum_{n=0}^{\infty}\frac{(-1)^{n+1}}{n!}\frac{\bigl(m^{2}(k)\bigr)^{n}}{\bigl(s-\frac{d}{2}+\frac{1}{2}+n\bigr)}.
\end{align}
%
Note that in principle, that the contribution of 
$I_{d}^{(3)}(s,k)$ is finite for odd dimensional space. 
Let us discuss the polar structure of $I_{d}^{(4)}(s,k)=\bigl(m^{2}(k)\bigr)^{\frac{d}{2}-s}\gamma\biggl(
s-\frac{d}{2},m^{2}(k)\biggr)$, where $\gamma(\alpha,x)=\int^{x}_{0}dt\,t^{\alpha-1}e^{-t}$ is an incomplete gamma function,
for $Re(\alpha)>0$. Using again a series representation for the incomplete gamma function we can write that
%
\begin{equation}
I_{d}^{(4)}(s,k)=
\sum_{n=0}^{\infty}\frac{(-1)^{n}}{n!}\frac{\bigl(m^{2}(k)\bigr)^{n}}{\bigl(s-\frac{d}{2}+n\bigr)}.
\end{equation}
\fi
the contribution of 
$I_{d}^{(4)}(s,k)$ is finite only for even dimensional space. Thus, it is not possible to define the Casimir-like energy per unit area associated to the Neumann Laplacian using an analytic regularization procedure in the Gaussian approximation \cite{dolan}. This obstruction is related to the presence of the zero mode \cite{2006}. 

With this in mind, we present the main result of this paper. For Dirichlet boundary conditions we can write $F_{d}(L)$  as 
\begin{equation}
F_{d}(L)=\sum_{k=k_{1}}^{k_{2}}c_{k}(a)
\exp\Biggl[\frac{k}{2}\zeta_{d}\biggl(-\frac{1}{2},k\biggr)\Biggr],
\label{quench}
\end{equation}
where $k_{1}=\lfloor\frac{m^{2}_{0}}{\sigma^{2}}\rfloor$ and 
$k_{2} =\lfloor\frac{1}{\sigma^{2}}
\bigl(\frac{1}{2L^{2}}+m_{0}^{2}\bigr)\rfloor$.

Examining the leading contribution of the series representation for the quenched free energy, where the correlation length of the fluctuations attains its maximum value, and with a suitable choice $a=\exp{\bigl(|\zeta_{d}\bigl(-\frac{1}{2},k_{1}\bigr)|\bigr)}$, we can write that the force per unit area is given by
\begin{equation}
f_{d}(L)=\frac{(-1)^{k_{1}+1}}{2k_{1}!}\frac{1}{\biggl(\prod_{i=1}^{d-1}L_{i}\biggr)}
\frac{\partial}{\partial L}\zeta_{d}\biggl(-\frac{1}{2},k_{1}\biggr).
\label{fdlend}
\end{equation}
Using the definition of $Z_{d}(s,k)$ we can rewrite (\ref{fdlend}) as follows, 
\begin{equation}
f_{d}(L)=\frac{(-1)^{k_{1}+1}}{2k_{1}!}
%\frac{1}{\biggl(\prod_{i=1}^{d-1}L_{i}\biggr)}
A(d)\frac{\partial}{\partial L}Z_{d}\biggl(-\frac{1}{2},k_{1}\biggr).
\end{equation}

We finally remark that this result that the fluctuation induced forces, attractive or repulsive, depends on  the strength of the disorder, as far as we know is new in the literature. This sign-changing of the fluctuation induced force should be testable in experiments~\cite{gambassi111, dotsenko111, gambassi112}.

\section{Conclusions} \label{sec:con}

Using the distributional zeta-function method, we discussed fluctuation-induced forces associated to a disordered Landau-Ginzburg model defined in a $d$-dimensional slab geometry. Assuming the Gaussian approximation in each moment of the partition function, we obtain a nearly critical scenario. For some specific strength of the disorder, the fluctuations associated to an order parameter-like quantity in a specific moment of the partition function becomes long-ranged. The induced-force per unit area in the case of Dirichlet boundary condition depends on the contribution coming  from the leading term, with the largest correlation length of the fluctuations. The sign of the induced-force depends on $\lfloor\frac{m^{2}_{0}}{\sigma^{2}}\rfloor$, being odd or even. A similar situation is obtained for a case with a dielectric surface and a permeable one, with  large dielectric constant $\epsilon$ and large permeability $\mu$ respectively. The transition between the attractive or repulsive behaviour depends on the ratio $\sqrt \frac{\mu}
{\epsilon}$. Our result remarking that the fluctuation induced forces, attractive or repulsive, depends on  the strength of the disorder is new in the literature.

To conclude we would like to point out that the quenched disorder generates fluctuations, which differs significantly from the thermal fluctuations. For pure, translational invariant systems with dimension of the order parameter being one, driven by thermal fluctuations, there is a unique temperature where the system becomes critical. For systems with quenched disorder, the correlation function associated to the order parameter remains long ranged for an enumerable set of values of the disorder. 

This led to the question of the 
the analytic structure of this disordered Landau-Ginzburg model. From the series representation of the average generating functional of connected correlation functions, one can obtain a series representation for the average generating functional of vertex functions. For a fixed disorder, i.e., $\sigma$ fixed, there is always a term in the series with $m_{0}^{2}-k\sigma ^{2}=0$. The argument follows: we define a sequence of critical $\sigma$ points,
this sequence has an accumulation point at $\sigma=0$. Notice that this occurs for any $m_{0}^{2}$. Therefore, we have infinitely many terms in the series that contributes with a divergent susceptibility, an infinite correlation length with power law decay of the correlation functions. The average generating functional of vertex functions has an infinite number of singularities.
In the complex $\sigma$ plane, this accumulation of singularities defines a natural boundary of analyticity, where there is no possibility of analytic extension \cite{landau,fro}. Actually, the limit $\sigma\rightarrow 0$ can not be achieved. The appearance of a natural boundary of a similar nature is studied in the prime number spectra in quantum field theory \cite{number111}.

%Referencias sobre a natural boundary;

A natural continuation of this work is to discuss the generalized Heisenberg ferromagnet with a $N$-dimensional order parameter \cite{biswas}, defined on a slab $\mathbb{R}^{d-1} \times [0,L]$,  invariant under the $O(N)$ symmetry group in the presence of quenched disorder. For $d>2$, where large correlation lengths may appear, one can discuss the fluctuation-induced force between boundaries in a Bose-Einstein condensate in the presence of disorder \cite{edery}. Finally, let us call the attention to the case where quantum and disorder-induced fluctuations are present in a system described by an Euclidean quantum $\lambda \phi ^{4}_{d+1}$  model with randomness. In that scenario a different situation is presented even in the three level approximation. This can be understood as the original system that present a non local term can be mapped into an statistical field theory model with spatial anisotropy~\cite{vojta1, vojta2, vojta3, vojta4}. These subjects are under investigation by the authors.

\begin{acknowledgements} 
 This work was partially supported by Conselho Nacional de Desenvolvimento Cient\'{\i}fico e Tecnol\'{o}gico - CNPq, the grant 
 - 301751/2019-6 (NFS). We thank the Engineering and Physical Sciences Research Council (EPSRC) (Grant No. EP/R513143/1 and EP/T517793/1) for financial
support (CDRC). We would like to thanks G. Krein and B. F. Svaiter for useful discussions.
\end{acknowledgements}

\bibliographystyle{apsrev4-1} % Tell bibtex which bibliography style to use

\bibliographystyle{apsrev4-1} % Tell bibtex which bibliography style to use
%\bibliography{xampl} % Tell bibtex which .bib file to use (this one is some example file in TexLive's file tree)

\end{document}